\documentstyle{mn}

\newif\ifAMStwofonts
\newread\epsffilein    % file to \read
\newif\ifepsffileok    % continue looking for the bounding box?
\newif\ifepsfbbfound   % success?
\newif\ifepsfverbose   % report what you're making?
\newdimen\epsfxsize    % horizontal size after scaling
\newdimen\epsfysize    % vertical size after scaling
\newdimen\epsftsize    % horizontal size before scaling
\newdimen\epsfrsize    % vertical size before scaling
\newdimen\epsftmp      % register for arithmetic manipulation
\newdimen\pspoints     % conversion factor
\def\epsfbox#1{\global\def\epsfllx{72}\global\def\epsflly{72}%
   \global\def\epsfurx{540}\global\def\epsfury{720}%
   \def\lbracket{[}\def\testit{#1}\ifx\testit\lbracket
   \let\next=\epsfgetlitbb\else\let\next=\epsfnormal\fi\next{#1}}%
\def\epsfgetlitbb#1#2 #3 #4 #5]#6{\epsfgrab #2 #3 #4 #5 .\\%
   \epsfsetgraph{#6}}%
\def\epsfnormal#1{\epsfgetbb{#1}\epsfsetgraph{#1}}%
\def\epsfgetbb#1{%
\openin\epsffilein=#1
\ifeof\epsffilein\errmessage{I couldn't open #1, will ignore it}\else
   {\epsffileoktrue \chardef\other=12
    \def\do##1{\catcode`##1=\other}\dospecials \catcode`\ =10
    \loop
       \read\epsffilein to \epsffileline
       \ifeof\epsffilein\epsffileokfalse\else
          \expandafter\epsfaux\epsffileline:. \\%
       \fi
   \ifepsffileok\repeat
   \ifepsfbbfound\else
    \ifepsfverbose\message{No bounding box comment in #1; using defaults}\fi\fi
   }\closein\epsffilein\fi}%
\def\epsfsetgraph#1{%
   \epsfrsize=\epsfury\pspoints
   \advance\epsfrsize by-\epsflly\pspoints
   \epsftsize=\epsfurx\pspoints
   \advance\epsftsize by-\epsfllx\pspoints
   \epsfxsize\epsfsize\epsftsize\epsfrsize
   \ifnum\epsfxsize=0 \ifnum\epsfysize=0
      \epsfxsize=\epsftsize \epsfysize=\epsfrsize
     \else\epsftmp=\epsftsize \divide\epsftmp\epsfrsize
       \epsfxsize=\epsfysize \multiply\epsfxsize\epsftmp
       \multiply\epsftmp\epsfrsize \advance\epsftsize-\epsftmp
       \epsftmp=\epsfysize
       \loop \advance\epsftsize\epsftsize \divide\epsftmp 2
       \ifnum\epsftmp>0
          \ifnum\epsftsize<\epsfrsize\else
             \advance\epsftsize-\epsfrsize \advance\epsfxsize\epsftmp \fi
       \repeat
     \fi
   \else\epsftmp=\epsfrsize \divide\epsftmp\epsftsize
     \epsfysize=\epsfxsize \multiply\epsfysize\epsftmp   
     \multiply\epsftmp\epsftsize \advance\epsfrsize-\epsftmp
     \epsftmp=\epsfxsize
     \loop \advance\epsfrsize\epsfrsize \divide\epsftmp 2
     \ifnum\epsftmp>0
        \ifnum\epsfrsize<\epsftsize\else
           \advance\epsfrsize-\epsftsize \advance\epsfysize\epsftmp \fi
     \repeat     
   \fi
   \ifepsfverbose\message{#1: width=\the\epsfxsize, height=\the\epsfysize}\fi
   \epsftmp=10\epsfxsize \divide\epsftmp\pspoints
   \vbox to\epsfysize{\vfil\hbox to\epsfxsize{%
      \includegraphics{#1}%
      \hfil}}%
\epsfxsize=0pt\epsfysize=0pt}%

{\catcode`\%=12 \global\let\epsfpercent=%\global\def\epsfbblit{%BoundingBox}}%
\long\def\epsfaux#1#2:#3\\{\ifx#1\epsfpercent
   \def\testit{#2}\ifx\testit\epsfbblit
      \epsfgrab #3 . . . \\%
      \epsffileokfalse
      \global\epsfbbfoundtrue
   \fi\else\ifx#1\par\else\epsffileokfalse\fi\fi}%
\def\epsfgrab #1 #2 #3 #4 #5\\{%
   \global\def\epsfllx{#1}\ifx\epsfllx\empty
      \epsfgrab #2 #3 #4 #5 .\\\else
   \global\def\epsflly{#2}%
   \global\def\epsfurx{#3}\global\def\epsfury{#4}\fi}%
\def\epsfsize#1#2{\epsfxsize}
\def\plotfiddle#1#2#3#4#5#6#7{\centering \leavevmode
    \vbox to#2{\rule{0pt}{#2}}
    \includegraphics{#1}}

\ifoldfss
  \ifCUPmtlplainloaded \else
    \NewTextAlphabet{textbfit} {cmbxti10} {}
    \NewTextAlphabet{textbfss} {cmssbx10} {}
    \NewMathAlphabet{mathbfit} {cmbxti10} {} % for math mode
    \NewMathAlphabet{mathbfss} {cmssbx10} {} %  "   "    "
  \fi
  \ifAMStwofonts
    \ifCUPmtlplainloaded \else
      \NewSymbolFont{upmath} {eurm10}
      \NewSymbolFont{AMSa} {msam10}
      \NewMathSymbol{\upi}     {0}{upmath}{19}
      \NewMathSymbol{\umu}     {0}{upmath}{16}
      \NewMathSymbol{\upartial}{0}{upmath}{40}
      \NewMathSymbol{\leqslant}{3}{AMSa}{36}
      \NewMathSymbol{\geqslant}{3}{AMSa}{3E}

      \let\leq=\leqslant \let\le=\leqslant
       
    \fi
  \fi
\fi % End of OFSS

\ifnfssone
  \newmathalphabet{\mathit}
  \addtoversion{normal}{\mathit}{cmr}{m}{it}
  \addtoversion{bold}{\mathit}{cmr}{bx}{it}
  \newmathalphabet{\mathbfit} % math mode version of \textbfit{..}
  \addtoversion{normal}{\mathbfit}{cmr}{bx}{it}
  \addtoversion{bold}{\mathbfit}{cmr}{bx}{it}
  \newmathalphabet{\mathbfss} % math mode version of \textbfss{..}
  \addtoversion{normal}{\mathbfss}{cmss}{bx}{n}
  \addtoversion{bold}{\mathbfss}{cmss}{bx}{n}
  \ifAMStwofonts
    \ifCUPmtlplainloaded \else
      \UseAMStwoboldmath
      \makeatletter
      \new@mathgroup\upmath@group
      \define@mathgroup\mv@normal\upmath@group{eur}{m}{n}
      \define@mathgroup\mv@bold\upmath@group{eur}{b}{n}
      \edef\UPM{\hexnumber\upmath@group}
      \new@mathgroup\amsa@group
      \define@mathgroup\mv@normal\amsa@group{msa}{m}{n}
      \define@mathgroup\mv@bold\amsa@group{msa}{m}{n}
      \edef\AMSa{\hexnumber\amsa@group}
      \makeatother
      \mathchardef\upi="0\UPM19
      \mathchardef\umu="0\UPM16
      \mathchardef\upartial="0\UPM40
      \mathchardef\leqslant="3\AMSa36
      \mathchardef\geqslant="3\AMSa3E

      \let\leq=\leqslant \let\le=\leqslant

    \fi
  \fi
\fi % End of NFSS release 1

\ifnfsstwo
  \DeclareMathAlphabet{\mathbfit}{OT1}{cmr}{bx}{it}
  \SetMathAlphabet\mathbfit{bold}{OT1}{cmr}{bx}{it}
  \DeclareMathAlphabet{\mathbfss}{OT1}{cmss}{bx}{n}
  \SetMathAlphabet\mathbfss{bold}{OT1}{cmss}{bx}{n}
  \ifAMStwofonts
    \ifCUPmtlplainloaded \else
      \DeclareSymbolFont{UPM}{U}{eur}{m}{n}
      \SetSymbolFont{UPM}{bold}{U}{eur}{b}{n}
      \DeclareSymbolFont{AMSa}{U}{msa}{m}{n}
      \DeclareMathSymbol{\upi}{0}{UPM}{"19}
      \DeclareMathSymbol{\umu}{0}{UPM}{"16}
      \DeclareMathSymbol{\upartial}{0}{UPM}{"40}
      \DeclareMathSymbol{\leqslant}{3}{AMSa}{"36}
      \DeclareMathSymbol{\geqslant}{3}{AMSa}{"3E}

      \let\leq=\leqslant \let\le=\leqslant

    \fi
  \fi
\fi % End of NFSS release 2

\ifCUPmtlplainloaded \else
  \ifAMStwofonts \else % If no AMS fonts
    \def\upi{\pi}
    \def\umu{\mu}
    \def\upartial{\partial}
  \fi
\fi

\title[The LCRS Galaxy-Galaxy Autocorrelation Function]
      {The Las Campanas Redshift Survey Galaxy-Galaxy Autocorrelation Function}
\author[D. L. Tucker et al.]
  {D.~L.~Tucker,$^{1,2}$ A.~Oemler,~Jr.,$^{3,4}$ R.~P.~Kirshner,$^5$ H.~Lin,$^6$ S.~A.~Shectman,$^3$ 
  \newauthor % starts a new line in the 
   S.~D.~Landy,$^3$ P.~L.~Schechter,$^7$ V.~M\"uller,$^2$ S.~Gottl\"ober,$^2$ and J.~Einasto$^8$\\
  $^1$Fermilab, 
      MS~127, 
      P.O.~Box~500, 
      Batavia, IL 60510 USA\\
  $^2$Astrophysikalisches Institut Potsdam, 
      An der Sternwarte 16, 
      D-14482 Potsdam, Germany\\
  $^3$Carnegie Observatories, 
      813 Santa Barbara Street, 
      Pasadena, CA 91101 USA\\
  $^4$Dept. of Astronomy, 
      Yale University, 
      New Haven, 
      CT 06520-8101 USA\\
  $^5$Harvard-Smithsonian Center for Astrophysics, 
      60 Garden Street,
      Cambridge, MA 02138 USA\\
  $^6$Dept. of Astronomy, 
      University of Toronto, 
      60 St. George St., 
      Toronto, ONT M5S 3H8, Canada\\
  $^7$Dept. of Physics 6-214, 
      Massachusetts Institute of Technology, 
      Cambridge, MA 02139 USA\\
  $^8$Tartu Astrophysical Observatory, 
      EE-2444 Toravere, Estonia}

\date{Accepted 1996 November 10. Received 1996 August 1}
\pagerange{\pageref{firstpage}--\pageref{lastpage}}
\pubyear{1996}

\begin{document}

\maketitle

\label{firstpage}

\begin{abstract}
Presented are measurements of the observed redshift-space
galaxy-galaxy autocorrelation function, $\xi_{\rm gg}(s)$, for the Las
Campanas Redshift Survey (LCRS).  For separations $2.0h^{-1}$~Mpc $ <
s < 16.4h^{-1}$~Mpc, $\xi_{\rm gg}(s)$ can be approximated by a power
law with slope $\gamma = -1.52 \pm 0.03$ and correlation length $s_0 =
6.28 \pm 0.27h^{-1}$~Mpc.  A zero-crossing occurs on scales of $\sim
30 - 40h^{-1}$~Mpc. On larger scales, $\xi_{\rm gg}(s)$ fluctuates
closely about zero, indicating a high level of uniformity in the
galaxy distribution on these scales.  In addition, two aspects of the
LCRS selection criteria -- a variable field-to-field galaxy sampling
rate and a 55~arcsec galaxy pair separation limit -- are tested and
found to have little impact on the measurement of $\xi_{\rm gg}(s)$.
Finally, the LCRS $\xi_{\rm gg}(s)$ is compared with those from
numerical simulations; it is concluded that, although the LCRS
$\xi_{\rm gg}(s)$ does not discriminate sharply among modern
cosmological models, redshift-space distortions in the LCRS $\xi_{\rm
gg}(s)$ will likely provide a strong test of theory.
\end{abstract}


\section{Introduction}

The original goals of the Las Campanas Redshift Survey (LCRS; Shectman
et al.\ 1996) were two-fold: firstly, to attempt to sample a `fair and
typical' volume of the nearby Universe in order to constrain the size
of the largest structures in the local galaxy distribution, and,
secondly, to use this sample to study galaxy clustering on a wide
variety of scales.  In order to accomplish these goals, it was decided
that the survey should be both spatially deep and angularly wide.
Indeed, the LCRS fulfills both these criteria in that it extends to a
redshift of $\sim 0.2$ and it is composed of a total of 6 alternating
$1\fdg5 \times 80\degr$ slices -- 3 each in the North and South
Galactic Caps.  A visual inspection of the slices indicates that the
largest walls and voids have sizes of $50 - 100h^{-1}$~Mpc, much
smaller than the largest survey dimensions, suggesting that the LCRS
does in fact encompass a fair sample.  Clearly, such data provide
exemplary material for the study of large-scale galaxy clustering.

Recently completed, the LCRS now contains 26,418 galaxy redshifts,
23,697 of which lie within the survey's official geometric and
photometric borders.  Accurate sky positions and Kron-Cousins $R$-band
photometry have come from CCD drift scans at the Las Campanas Swope
1-m telescope; spectra have been obtained at the Las Campanas Du~Pont
2.5-m telescope, originally with a 50-fibre Multi-Object Spectrograph
(MOS) and later with a 112-fibre MOS.  For observing efficiency, all
the fibres are used, but each MOS field is observed only once.  Hence,
the LCRS is a collection of 50-fibre fields (with nominal apparent
magnitude limits of $16.0 \leq R < 17.3$) and 112-fibre fields (with
nominal apparent magnitude limits of $15.0 \leq R < 17.7$).  Thus,
selection criteria vary from field to field, but they are carefully
documented and are therefore easily taken into account.  Observing
each field only once, however, creates an additional selection effect:
the individual fibres' protective tubing prevents the observation of
galaxy pairs within 55~arcsec of each other.  Hence, the cores of rich
clusters may be undersampled, potentially causing underestimates in
measurements of small-scale galaxy clustering.  In this paper, we will
consider the LCRS redshift-space galaxy-galaxy autocorrelation
function, $\xi_{\rm gg}(s)$.  We will also examine the influence on it
from the fibre-separation limit and the field-to-field sampling
variations.  Finally, we will compare the LCRS $\xi_{\rm gg}(s)$ with
those derived from numerical simulations.

\section{Method}

To account for the survey geometry and for the field-to-field
variations in the nominal apparent magnitude limits and in the
sampling fraction, we generate catalogues of random galaxies over the
same survey volume with the same field-to-field characteristics as the
observed LCRS catalogue.  (Note that, although its inclusion would
tend to diminish the effects of undersampling on small scales, for
ease of interpretation the fibre-separation limit is {\it not \/}
implemented into the random catalogues.)  The redshift distribution of
random galaxies is determined via the LCRS luminosity function
described in Lin et al.\ (1996a), incorporating the subtleties
detailed in Section~3.2 of that paper.  We then calculate $\xi_{\rm
gg}(s)$ according to the Hamilton (1993) formalism,
\begin{equation}
1 + \xi_{\rm gg}(s) =  \frac{RR(s)}{DR(s)} \times \frac{DD(s)}{DR(s)}, 
\end{equation}
where $DD(s)$, $DR(s)$, and $RR(s)$ are, respectively, the weighted
data-data pair count ($\Sigma_{i \neq j} w_i w_j$), the weighted
data-random pair count ($\Sigma w_i w^{\rm r}_j$), and the weighted
random-random pair count ($\Sigma_{i \neq j} w^{\rm r}_i w^{\rm r}_j$)
for the (comoving) separation $s$; $RR/DR$ is a measure of the
relative mean density of galaxies in the observed and random
catalogues.  Aside from small differences in the large-scale
normalisation, our results change little if a classic Davis \& Peebles
(1983) approach is employed (see Fig.~1 below).  We prefer the Hamilton
formalism, however, since it is less affected by uncertainties in the
mean density of galaxies (see also Landy \& Szalay 1993).

For the pair counts, Hamilton (1993) suggests the minimum variance
weight for each galaxy, 
\begin{equation}
w_i  = \frac{1}{ 1 + 4 \upi n^{\rm exp}_{\rm fld}(z_i) J_3(s) }
\end{equation}
(likewise for $w^{\rm r}_i$), where $n^{\rm exp}_{\rm fld}(z_i)$ is
the number density of galaxies one would expect to observe at a
redshift $z_i$ for a given field under the constraints of the
luminosity function and of the selection effects peculiar to that
field, and where
\begin{equation}
J_3(s) \equiv \int_0^s x^2 \xi(x) dx.
\end{equation}
For this integral, we have approximated $\xi$ by a power law with
slope $-1.6$ and correlation length $s_0 = 6.0h^{-1}$~Mpc, which is a
good approximation for separations $2h^{-1}~\mbox{Mpc} \la s \la
20h^{-1}~\mbox{Mpc}$ (see Sec.~3.1).  Results are robust for
reasonable values of the slope and correlation length.

Due to the typical distances involved, galaxy velocities are corrected
for CMB-dipole motion (Lineweaver et al.\ 1996) rather than for a
Virgo infall model; tests indicate, however, that simply using
heliocentric velocities has little effect on the analysis.  Positions
are then converted into comoving distances assuming $H_0
=100h$~km~s$^{-1}$~Mpc$^{-1}$, $q_0 = 0.5$, and $\Lambda = 0$.  In
order to avoid the weights `blowing up' at the survey's extremal
distances, we confine our analysis of $\xi_{\rm gg}(s)$ to those
galaxies in the LCRS with velocities of $10,000$~km~s$^{-1} \leq
cz_{\rm CMB} \leq 45,000$~km~s$^{-1}$ and with $R$-band absolute
magnitudes of $-22.50 \leq M_R - 5\log h \leq -18.50$; 19,314 galaxies
from the official LCRS sample meet these criteria.  The random
catalogues typically contain a similar number of galaxies.

In what follows, $1\sigma$ error bars are estimated by calculating
$\xi_{\rm gg}(s)$ in independent subregions of the LCRS and taking the
standard deviation of the mean.  For $s \leq 200h^{-1}$~Mpc, the LCRS
is split into four separate subregions (Northern Galactic Cap data, RA
$\leq 12^{\rm h}~42^{\rm m}$ and RA $> 12^{\rm h}~42^{\rm m}$;
Southern Galactic Cap, RA $\leq 00^{\rm h}~42^{\rm m}$ and RA $>
00^{\rm h}~42^{\rm m}$).  For $s > 200h^{-1}$~Mpc, only two
independent subregions are considered -- the Northern Galactic Cap and
the Southern Galactic Cap survey volumes.  [Note: preliminary tests
indicate that, for our sample, the bootstrap errors (Ling, Frenk, \&
Barrow 1986) are comparable in magnitude to those calculated with the
above method.]

\section{Results}

\subsection{The Observed LCRS $\xi_{\rm gg}(s)$}  

\begin{figure}
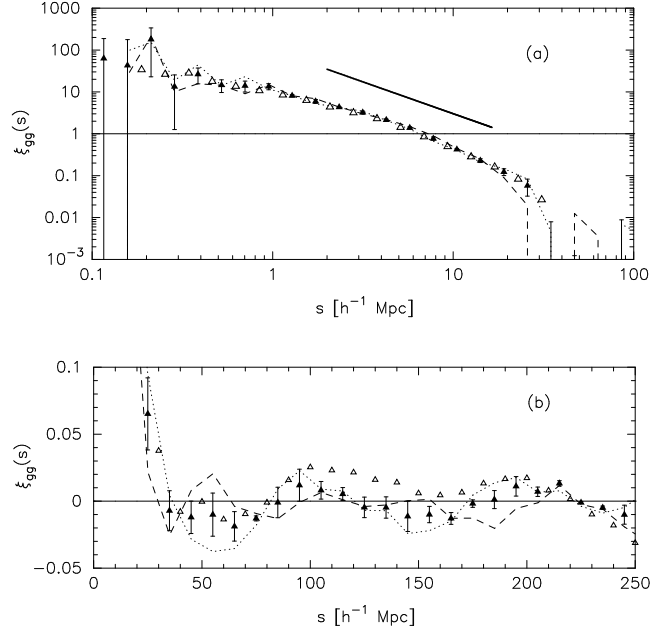

\plotfiddle{xipaper4mn-fig1a.ps}{5.00in}{-90}{45}{45}{-182}{455}
\plotfiddle{xipaper4mn-fig1b.ps}{5.00in}{-90}{45}{45}{-180}{700}
\vspace{-17.50cm} 
\caption{The observed LCRS $\xi_{\rm gg}(s)$ for (a) small-to-intermediate 
scales and for (b) intermediate-to-large scales.  The {\em filled
triangles\/} denote $\xi_{\rm gg}(s)$ for the combined North and South
Galactic Cap sample, the {\em dashed line\/} for the Northern Cap
sample alone, and the {\em dotted line\/} for the Southern Cap sample
alone; these measurements all use the Hamilton formalism. For
comparison, the LCRS $\xi_{\rm gg}(s)$ calculated using the Davis \&
Peebles formalism is also presented ({\em unfilled triangles}); here,
as with the Hamilton formalism, the weighting scheme of equation (2)
is employed.  For clarity, only the {\em filled triangles\/} show
error bars.  Finally, a $-1.52$ power law, offset, is shown in (a) for
the interval $2.0h^{-1}$~Mpc $ < s < 16.4h^{-1}$~Mpc ({\em thick solid
line}).}
\end{figure}

The results for the observed LCRS $\xi_{\rm gg}(s)$ can be found in
Figure~1.  For separations $2.0h^{-1}$~Mpc $ < s < 16.4h^{-1}$~Mpc, we
find that the observed LCRS $\xi_{\rm gg}(s)$ can be approximated by a
power law with slope $\gamma = -1.52 \pm 0.03$ and correlation length
$s_0 = 6.28 \pm 0.27h^{-1}$~Mpc (Fig.~1a, Hamilton formalism).  A
zero-crossing occurs at $s \sim 30 - 40h^{-1}$~Mpc.  On larger scales,
$\xi_{\rm gg}(s)$ fluctuates closely about zero, evidence of a high
level of uniformity in the galaxy distribution on these scales
(Fig.~1b).  Although possible small-amplitude ($\delta \xi_{\rm gg}
\approx 0.01 \pm 0.01$) secondary maxima do appear at $\sim
100h^{-1}$~Mpc and at $\sim 200h^{-1}$~Mpc, whether these (and other,
larger-scale) relative extrema are characteristic of the galaxy
distribution itself or merely statistical fluctuations within the
sampled volume remains a matter of debate and is still under
investigation.  It is interesting to note, however, that these
large-scale features -- and the differences between the LCRS North and
South Galactic Cap samples -- are reflected in Landy et al.\ (1996)'s
determination of the LCRS 2D power spectrum (their Fig.~2), and that
Doroshkevich et al.\ (1997)'s core-sampling analysis of the LCRS
reveals a scale of $\sim 100h^{-1}$~Mpc for the typical separation of
sheet- or wall-like structures.  Furthermore, in an independent
sample, Einasto et al.\ (1997) find similar features in the
autocorrelation function for clusters in the environments of rich
superclusters.

%We also note that the differences in the LCRS $\xi_{\rm gg}(s)$
%between the North and the South Galactic Cap regions -- in particular, the
%earlier break from a power law for the LCRS Northern Cap $\xi_{\rm
%gg}(s)$ -- may at least partially explain the disagreement between
%measurements of the {\em angular \/} autocorrelation function,
%$\omega_{\rm gg}(\theta)$, for the northern-sky Shane-Wirtanen
%catalogue (Groth \& Peebles 1977) and the southern-sky APM catalogue
%(Maddox et al.\ 1990a): perhaps the disagreement between the two
%measurements of $\omega_{\rm gg}(\theta)$ may be more a reflection of
%small-but-real differences in the amount of clustering observed in the
%two galactic hemispheres rather than of subtle systematic errors in
%either of the two analyses.

Finally, we compare the LCRS $\xi_{\rm gg}(s)$ with those from two
other modern redshift surveys (Fig.~2): Park et al.'s (1994)
determination for a sample of the extended CfA Redshift Survey (CfA2)
and Loveday et al.'s (1996) determination for the Stromlo-APM Redshift
Survey.  The Park et al.\ sample under discussion covers 2~sr of sky
out to a comoving depth of $101h^{-1}$~Mpc and consists of 7453
galaxies with $m_{B(0)} \le 15.5$ (their CfA101m sample).  The
Stromlo-APM Redshift Survey, the basis of the Loveday et al.\
analysis, covers 1.3~sr of sky and contains 1787 galaxies with $b_J
\le 17.15$ which were randomly selected at a rate of 1 in 20 from the
APM galaxy catalogue (Maddox et al.\ 1990).  The autocorrelation
functions from both these surveys match the LCRS $\xi_{\rm gg}(s)$
quite well on small scales ($s \la 10h^{-1}$~Mpc).  Unfortunately, the
CfA2 $\xi_{\rm gg}(s)$, calculated using the Davis \& Peebles
formalism becomes unreliable on scales $s \ga 15h^{-1}$~Mpc, due to
the fractional uncertainty in the mean number density of galaxies
(Park et al.\ 1994).  On the other hand, the Stromlo-APM $\xi_{\rm
gg}(s)$, calculated via the Hamilton formalism, is less hampered by
uncertainties in the mean galaxy number density; that it appears to
show some excess power relative to the LCRS $\xi_{\rm gg}(s)$ on large
scales ($s \ga 20 - 30h^{-1}$~Mpc) may indicate real differences in
the clustering properties of the LCRS and the Stromlo-APM samples.

\begin{figure}
\plotfiddle{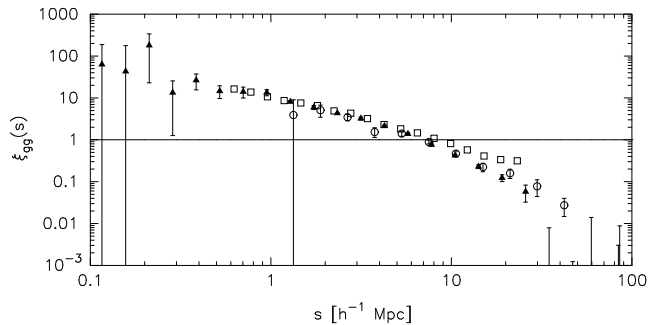}{5.00in}{-90}{45}{45}{-182}{455}
\vspace{-8.50cm} 
\caption{A comparison of the LCRS $\xi_{\rm gg}(s)$ (Hamilton formalism)
from Fig.~1 ({\em filled triangles\/}) with determinations from two
other optically selected modern surveys: Park et al.'s (1994) minimum
variance estimate of $\xi_{\rm gg}(s)$ (Davis \& Peebles formalism)
for the CfA Redshift Survey ({\em unfilled squares, no error bars
available\/}) and Loveday et al.'s (1996) minimum variance estimate of
$\xi_{\rm gg}(s)$ (Hamilton formalism) for the Stromlo-APM Redshift
Survey ({\em unfilled circles with error bars\/}).}
\end{figure}

\subsection{The Fibre Separation Test}

\begin{figure}
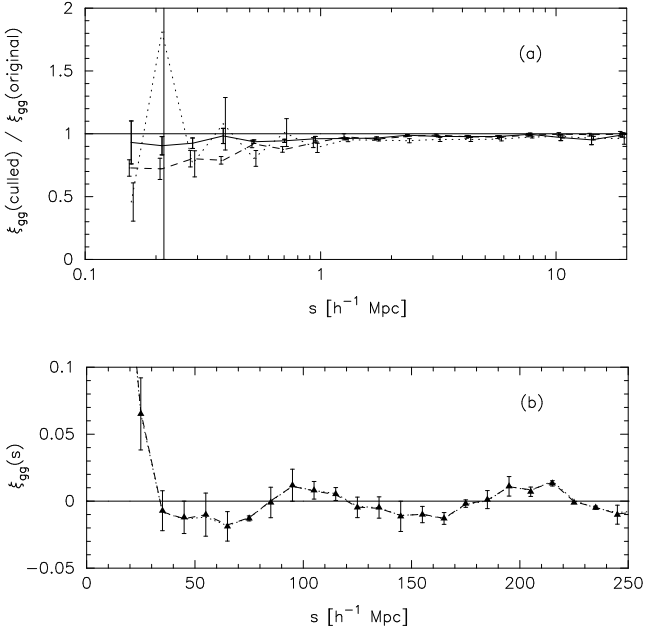

\plotfiddle{xipaper4mn-fig3a.ps}{5.00in}{-90}{45}{45}{-182}{455}
\plotfiddle{xipaper4mn-fig3b.ps}{5.00in}{-90}{45}{45}{-180}{700}
\vspace{-17.50cm} 
\caption{The Fibre Separation Test. (a) On small-to-intermediate
scales we plot the ratio of $\xi_{\rm gg}(s)$ of a catalogue culled to
a given fibre separation limit to that of the original catalogue,
$\xi_{\rm gg}({\rm culled}) / \xi_{\rm gg}({\rm original})$. This
ratio is plotted only on those scales where both $\xi_{\rm gg}({\rm
culled})$ and $\xi_{\rm gg}({\rm original})$ are clearly positive.
For the LCRS data, $\xi_{\rm gg}({\rm original})$ is the LCRS
$\xi_{\rm gg}(s)$ from Fig.~1a (Hamilton formalism); we plot this
ratio for the LCRS catalogue culled to 90~arcsec ({\em dashed line\/})
and to 120~arcsec ({\em dotted line\/}) fibre separation limits.
Error bars are the standard deviation of the mean of $\xi_{\rm
gg}({\rm culled}) / \xi_{\rm gg}({\rm original})$ from the same four
subvolumes used to calculate the error in the LCRS $\xi_{\rm gg}(s)$
on these scales.  For the SCDM mock slices, $\xi_{\rm gg}({\rm
original})$ is $\xi_{\rm gg}(s)$ with no fibre separation limit;
$\xi_{\rm gg}({\rm culled})$ is $\xi_{\rm gg}(s)$ for the mock slice
culled to a fibre separation limit of 55~arcsec.  We plot the mean of
this ratio based upon 5 individual mock slices ({\em solid line\/}),
where the error bars are the standard deviation of the mean.  (b) On
intermediate-to-large scales, we simply plot $\xi_{\rm gg}(s)$: the
{\em filled triangles\/} are as in Fig.~1, the {\em dashed line\/} is
the LCRS $\xi_{\rm gg}(s)$ with a 90~arcsec fibre separation limit,
and the {\em dotted line\/} is the LCRS $\xi_{\rm gg}(s)$ with a
120~arcsec fibre separation limit.}
\end{figure}

To test the effects of fibre separation, we have artificially
increased the fibre separation criterion in each MOS field from the
original 55~arcsec -- first to 90~arcsec and then to 120~arcsec.  This
was done by culling the original LCRS catalogue -- field-by-field --
of one galaxy in any pair separated on the sky by less than the
artificial fibre limit.  We can then extrapolate backwards to estimate
how the original 55~arcsec separation limit affects the measurement of
$\xi_{\rm gg}(s)$.  For a more direct test, we have also taken 5
individual mock slices -- each based upon the Standard Cold Dark
Matter (SCDM) simulations of Section~3.4 -- and then imposed a
55~arcsec fibre separation limit in a manner akin to that for a real
LCRS slice.

The effect is negligible on all but the smallest scales, $s \la
1h^{-1}$~Mpc (Fig.~3).  At these scales, however, Poisson errors begin
to dominate the results, making it difficult to assess the exact
magnitude of the effect.  Nonetheless, even at separations of $s \sim
0.3h^{-1}$~Mpc, it appears that the LCRS 55~arcsec fibre separation
limit results only in a $\sim 10 - 20$ per cent underestimate of the
`true' $\xi_{\rm gg}(s)$.  On scales of $1h^{-1}$~Mpc $\la s \la
20h^{-1}$~Mpc, measurements of $\xi_{\rm gg}(s)$ are depressed
typically by $\la 5$ per cent due to fibre separation, and, on the
largest scales (Fig.~3b), the fibre separation effect is hardly
evident.

\subsection{The Uniform Sampling Test}

\begin{figure}
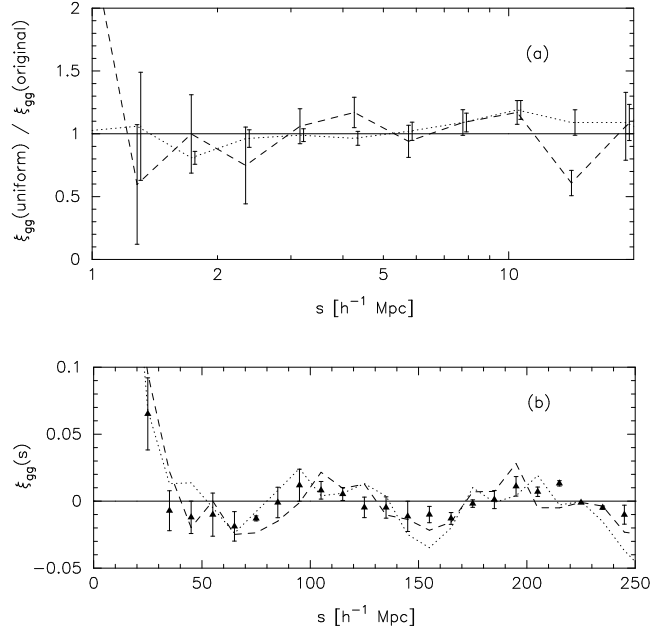

\plotfiddle{xipaper4mn-fig4a.ps}{5.00in}{-90}{45}{45}{-182}{455}
\plotfiddle{xipaper4mn-fig4b.ps}{5.00in}{-90}{45}{45}{-180}{700}
\vspace{-17.50cm} 
\caption{The Uniform Sampling Test.  (a) On small-to-intermediate
scales we plot the ratio of $\xi_{\rm gg}$ for the LCRS catalogue
culled to uniform sampling to $\xi_{\rm gg}(s)$ for the full LCRS
catalogue (such as in Fig.~1).  This ratio is plotted only on those
scales where $\xi_{\rm gg}(s)$ is clearly positive for both the
original and the uniform catalogues.  The ({\em dashed line\/})
denotes the ratio for the uLCRS catalogue, and the ({\em dotted line})
for the uLCRS112 catalogue.  Error bars are the standard deviation of
the mean of $\xi_{\rm gg}({\rm uniform}) / \xi_{\rm gg}({\rm
original})$ from the same four subvolumes used to calculate the error
in the LCRS $\xi_{\rm gg}(s)$ on these scales.  (b) On
intermediate-to-large scales, the {\em filled triangles\/} are as in
Fig.~1, the {\em dashed line\/} denotes the $\xi_{\rm gg}(s)$ from the
uniformly sampled uLCRS catalogue, and the {\em dotted line\/} denotes
the $\xi_{\rm gg}(s)$ from the uniformly sampled uLCRS112 catalogue.}
\end{figure}

Although the variations in the field-to-field sampling are included
both in the random catalogue of galaxies and in the galaxy weighting
[equation~(2)], there may be concern that these variations may still
result in some residual aliasing of power in the observed $\xi_{\rm
gg}(s)$ [this would be in addition to any aliasing due to the slice
geometry of the Survey; cf. Kaiser \& Peacock (1991)].  Therefore, to
test the effects of the field-to-field variations in the nominal
apparent magnitude limits and in the sampling fraction, two uniform
galaxy catalogues were extracted from the LCRS: a catalogue of uniform
field sampling fraction (21 per cent, the most restrictive in the
LCRS) and of uniform apparent magnitude limits ($16.33 \leq R <
17.13$) was extracted from the full 50- \& 112-fibre LCRS.  This
uniform catalogue (the uLCRS) contains $\sim 1/7$ the number of
galaxies from the original catalogue.  A second uniform catalogue
(containing $\sim 1/3$ the number galaxies of the original) was
extracted from just the set of 112-fibre fields.  This catalogue (the
uLCRS112) has a uniform field sampling fraction of 34 per cent (the
most restrictive from the 112-fibre fields) and uniform apparent
magnitude limits, $15.18 \leq R < 17.47$.

Owing to the greater statistical noise of the two uniform catalogues
(due to their greatly truncated samples), it is hard to distinguish
any significant difference between the $\xi_{\rm gg}(s)$ derived from
the full LCRS sample and those from the uLCRS and the uLCRS112
(Fig.~4).  Indeed, on small-to-intermediate scales (Fig.~4a), any
effect must be of a magnitude of $\la 20$ per cent; on large scales
(Fig.~4b), any effect must be of amplitude $\delta\xi_{\rm gg} \la
0.01$.  The present analysis is consistent with the null hypothesis
that the LCRS $\xi_{\rm gg}(s)$, as calculated, does {\em not\/}
suffer from aliasing due to field-to-field sampling variations.  The
placement of more stringent limits on this effect will require
extensive testing with mock catalogues covering the full LCRS volume.

\subsection{Comparison with Theory}

\begin{figure}
\plotfiddle{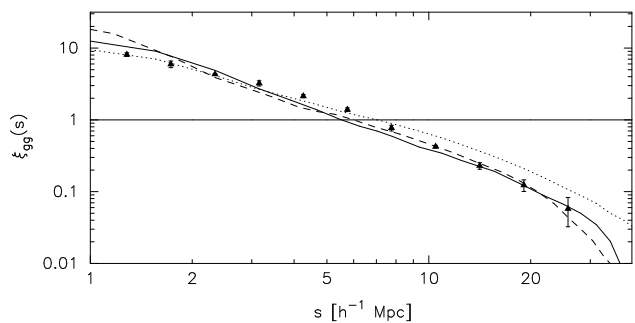}{5.00in}{-90}{45}{45}{-182}{455}
\vspace{-8.50cm} 
\caption{Comparison with theory at small-to-intermediate scales.  
The {\em filled triangles\/} are as in Fig.~1; the {\em solid line\/}
denotes the $\xi_{\rm gg}(s)$ from the SCDM model, the {\em dashed
line\/} denotes the $\xi_{\rm gg}(s)$ from the $\Lambda$CDM model, and
the {\em dotted line\/} denotes the $\xi_{\rm gg}(s)$ from the BSI
model.}
\end{figure}

For this study, galaxy catalogues for three cosmological models were
produced -- one based on the Standard Cold Dark Matter (SCDM) model
($H_0 = 50$~km~s$^{-1}$~Mpc$^{-1}$, $\Omega_0 = 1$), another on a
non-zero cosmological constant model ($\Lambda$CDM; $H_0 =
70$~km~s$^{-1}$~Mpc$^{-1}$, $\Omega_0 = 0.35$, $\Omega_{\Lambda} =
0.65$), and a third on the Broken Scale Invariance (BSI) model
predicted by a double--inflation scenario (Gottl\"ober, M\"uller, \&
Starobinsky 1991; Gottl\"ober, M\"ucket, \& Starobinsky 1994).  All
models are COBE normalised.
%; so we expect slightly antibiased galaxy
%formation in the SCDM model and a biasing factor of $b \sim 1.5 - 2$
%for the other two models.  
We have studied structure formation in $200h^{-1}$~Mpc and
$150h^{-1}$~Mpc boxes for the SCDM/BSI and the $\Lambda$CDM
simulations, respectively, using a PM code with $128^3$ particles in
$256^3$ grid cells (Kates et al.\ 1995).

Galaxies are identified as local maxima in the density field of dark
matter, with an overdensity of 30 for the BSI and $\Lambda$CDM
simulations and an overdensity of 70 for the more evolved SCDM.  These
thresholds are a compromise between the results of spherical collapse
estimates and the low spatial resolution of the simulation.  Then we
collected all the particles into `halos' with a diameter of a grid
cell length and produced a random realisation of the LCRS luminosity
function.  To this end, we used mass-to-light ratios of $M/L =$ 80,
150, and 240 for the $\Lambda$CDM, BSI, and SCDM simulations,
respectively, in order to provide the total luminosity predicted by
the integrated LCRS luminosity function for the summed masses of the
peak-selected halos. Further, we distributed the `galaxies' within the
volume of the halo with a Gaussian spatial displacement and with a
Gaussian velocity dispersion with a variance according to the virial
theorem.  In this way, we correct for the low value of the small-scale
velocity dispersion characteristic of PM simulations.  The resulting
galaxy catalogue has a wide assortment of galaxy groups and clusters,
but it undersamples field galaxies.

In Figure~5, we compare the redshift-space autocorrelation functions
of these simulated galaxy catalogues with the results from the LCRS.
All three models well represent the observations.  In particular, over
the range $1 - 30 h^{-1}$~Mpc, we obtain a $\xi_{\rm gg}(s)$ with the
quite gentle slope of $\gamma \simeq -1.5$, comparable to that of the
observed sample.  Even so, we note that the small-scale peculiar
velocities of the galaxy tracers in the SCDM simulation are higher
than in either of the other two models.  Therefore, the redshift-space
distortions in the LCRS $\xi_{\rm gg}(s)$ promise to be a strong
discriminating test of dark matter models (Lin et al.\ 1996b).

Looking more closely at the data we see that both the $\Lambda$CDM and
the BSI autocorrelation functions show, on some scales, excess
clustering relative to that of the LCRS -- at small scales for the
$\Lambda$CDM model and in the linear regime [$\xi_{\rm gg}(s) < 1$]
for BSI.  Both effects are influenced by the procedure of the galaxy
selection, i.e., by the effective biasing of dark and luminous matter.
We can thus conclude that the high quality of the LCRS $\xi_{\rm
gg}(s)$ is a sensitive tool for verifying the modelling of this
physically complex process, which could only be done crudely in our
present simulations.

\section{Conclusions}

We have presented a detailed study of the LCRS redshift-space
galaxy-galaxy autocorrelation function, $\xi_{\rm gg}(s)$.  We found
that the observed LCRS $\xi_{\rm gg}(s)$ can be approximated by a
power law with slope $\gamma = -1.52 \pm 0.03$ and correlation length
$s_0 = 6.28 \pm 0.27h^{-1}$~Mpc for separations $2.0h^{-1}$~Mpc $ < s
< 16.4h^{-1}$~Mpc, and that $\xi_{\rm gg}(s)$ first drops below zero
at $s \sim 30 - 40h^{-1}$~Mpc.  For $s \ga 50h^{-1}$~Mpc, $\xi_{\rm
gg}(s)$ fluctuates closely about zero, indicating that the galaxy
distribution is quite homogeneous on these scales.  Although possible
secondary maxima, of amplitude $\delta \xi_{\rm gg} \approx 0.01 \pm
0.01$, are observed at $\sim 100h^{-1}$~Mpc and $\sim200h^{-1}$~Mpc,
their true significance is still under investigation.

Two selection criteria peculiar to fibre-optic redshift surveys --
field-to-field variations in the rate of galaxy sampling and a limit
to the observation of close galaxy pairs due to fibre-positioning
constraints -- were found, on most scales ($s \ga 1h^{-1}$~Mpc), to
have little or no effect on the measurement of the observed $\xi_{\rm
gg}(s)$.  On the smallest scales ($s \la 1h^{-1}$~Mpc), the 55~arcsec
fibre separation limit increasingly dampens the magnitude of the
observed $\xi_{\rm gg}(s)$, by about 5 per cent at $s \approx
1h^{-1}$~Mpc to $10 - 20$ per cent at $s \approx 0.3h^{-1}$~Mpc to
perhaps $\ga 50$ per cent for $s \la 0.15h^{-1}$~Mpc.

Finally, the observed LCRS $\xi_{\rm gg}(s)$ was compared with those
from numerical simulations of the SCDM, $\Lambda$CDM, and BSI models.
Although all of the model autocorrelation functions fit the
observations reasonably well, we anticipate that the redshift-space
distortions in the LCRS $\xi_{\rm gg}(s)$ will prove to be an even
stronger test of the models.  Furthermore, we conclude that, due to
its high quality, the LCRS $\xi_{\rm gg}(s)$ provides a sensitive
means for discriminating among the galaxy identification procedures of
N-body simulations.

\section*{Acknowledgments}

The Las Campanas Redshift Survey has been supported by NSF grants AST
87-17207, AST 89-21326, and AST 92-20460. HL also acknowledges support
from NASA grant NGT-51093.  Thanks also to Dr.\ C. Heller
(G\"ottingen) and to the anonymous referee for their respective
suggestions at the inception and the completion of this work.

\bsp

\label{lastpage}

\end{document}